\documentclass{article}

\usepackage{arxiv}
\usepackage[numbers,sort&compress]{natbib}

\usepackage[utf8]{inputenc} % allow utf-8 input
\usepackage[T1]{fontenc}    % use 8-bit T1 fonts
\usepackage{hyperref}       % hyperlinks
\usepackage{url}            % simple URL typesetting
\usepackage{booktabs}       % professional-quality tables
\usepackage{amsfonts}       % blackboard math symbols
\usepackage{nicefrac}       % compact symbols for 1/2, etc.
\usepackage{microtype}      % microtypography
\usepackage[ruled,vlined]{algorithm2e} % for algorithm environment
\usepackage{float}          % for [H] float placement
\usepackage{lipsum}		% Can be removed after putting your text content
\usepackage{graphicx}
\usepackage{doi}
\usepackage{listings}
\usepackage[section]{placeins}

\title{Development of a Programming Based Kinetic Model for Two Stage Composting of Solid Waste}

\author{
  Zarif Tanzim Aziz\thanks{Corresponding author: zarif.ta.82@gmail.com} \\
  Department of Civil Engineering\\
  Khulna University of Engineering \& Technology\\
  Khulna, Bangladesh \\
  \texttt{zarif.ta.82@gmail.com} \\
  \And
  Md. Mahfuzur Rahman \\
  Department of Computer Science and Engineering\\
  Ahsanullah University of Science and Technology\\
  Dhaka, Bangladesh \\
  \texttt{} \\
  \And
  Quazi Hamidul Bari \\
  Department of Civil Engineering\\
  Khulna University of Engineering \& Technology\\
  Khulna, Bangladesh \\
  \texttt{} \\
}

\renewcommand{\headeright}{}
\renewcommand{\undertitle}{}
\renewcommand{\shorttitle}{Development of a Programming Based Kinetic Model for Two Stage Composting of Solid Waste}

\hypersetup{
pdftitle={Development of a Programming Based Kinetic Model for Two Stage Composting of Solid Waste},
pdfsubject={Waste management, composting, kinetic modelling},
pdfauthor={Zarif Tanzim Aziz, Md. Mahfuzur Rahman, Quazi Hamidul Bari},
pdfkeywords={waste management, kinetic model, aeration, programming, composting},
}

\begin{document}
\maketitle

% --- Abstract ---
\begin{abstract}
As the world is moving toward sustainable development, there is an important need to adopt sustainable waste management solutions of biodegradable solid waste, such as composting, which offers significant advantages over traditional methods like landfilling and incineration by reducing greenhouse gas emissions, enriching soil fertility, and minimizing landfill waste. However, optimizing the composting process governed by factors like aeration, moisture, and carbon-to-nitrogen ratio often relies on complex mathematical models that are difficult to interpret and apply. To overcome this challenge, a user-friendly programming-based two-stage kinetic model has been developed which simplifies composting efficiency analysis, with the first stage covering the initial 28 days of decomposition and the second stage evaluating further degradation, making the process more accessible and actionable for sustainable waste management.

\end{abstract}

% ============================================================
\section{Introduction}

Each year, approximately 38 billion metric tons of solid waste are produced all over the world \citep{ref1}. Bangladesh's urban regions generate 23,688 tons of trash every day \citep{ref2}. In developing countries, most of the solid waste is dumped in open landfills or burned in the open air \citep{ref3, ref4}. These methods are not environmentally friendly and are not permitted. Because of open dumping, landfilling contaminates soil and water due to the production of leachates. It also causes air pollution as it produces carbon dioxide and other gases \citep{ref5, ref6}. % [rev] "methane," deleted from this list per the revision.
These solid wastes contain metals, organic compounds, micropollutants, inorganic macro-components, xenobiotic compounds, and dioxins, which affect water, soil, air, and human health \citep{ref7}. It is high time we needed a better, more economical, and environmentally friendly process to deal with solid waste.

There are mainly two portions of solid waste: the organic portion and the inorganic portion. More than 70\% of the municipal solid waste generated in Bangladesh contains organic material \citep{ref2}. Among them are putrescible organic waste (vegetable waste, paper waste) and non-putrescible organic waste (plastic, rubber). The putrescible organic waste degrades in natural conditions. It takes a long time to fully degrade under normal environmental conditions. While degrading, various types of microorganisms and bacteria grow inside it, producing different gases like methane, and a bad odor comes out of it. But under a controlled environment, time can be reduced by up to 4 to 10 weeks by composting \citep{ref8}. And it can reduce weight and volume significantly \citep{ref9}. If we can control the temperature, moisture content, C/N ratio, the rate of air flow, and some other parameters, the degradation of organic waste can be done faster \citep{ref10}. If not properly managed, the biodegradable component of organic solid waste will decompose and release unpleasant odors into the environment \citep{ref11}. Composting is the degradation of highly concentrated organic wastes in the presence of oxygen (an aerobic condition) to carbon dioxide and water, whereby the biologically generated heat is sufficient to raise the temperature of the composting mass to the thermophilic range (50\textdegree{}C to 65\textdegree{}C) \citep{ref12}. It is a suitable approach for managing the organic portion of solid waste \citep{ref13}.

Composting is the gradual degradation of biodegradable organic wastes in the presence of oxygen and microorganisms into a stable, humus-like material known as compost. The composting process generally produces heat \citep{ref12}. Composting is useful for manure handling because it destroys pathogens, can be used as a soil conditioner, and lowers the risk of pollution \citep{ref9}.

There are few mathematical models for calculating composting parameters. Mathematical models assist in analyzing the energy balance by taking into account aspects such as decomposition degree and heat loss. Maintaining proper temperatures improves compost quality and speeds up the process \citep{ref14}.

% [rev] "experimental physical model" inserted per the revision.
For calculating the compostability of biodegradable organic waste, a model has been established by Bari and Koenig (2012). The cost of an experimental physical model of composting is very high and time-consuming. Using the mathematical kinetic model of composting biodegradable organic waste, the composting parameter can be easily calculated \citep{ref15}. A Microsoft Excel file has already been generated to calculate the different parameters of composting. But there are many drawbacks to a mathematical model developed using Microsoft Excel. It is not user-friendly; the formula inserted in the cell of the Excel file can be mistakenly changed by the user without noticing. Once the file has been corrupted, it will be quite impossible to find the error inside it. So, a programming-based model is needed. The second stage of composting is also done to see further degradation in this model.

A programming language, Python, was used for the developed programming-based kinetic model. Python, for example, can handle large datasets and complex calculations more efficiently than Excel. As the model grows in complexity, programming languages provide better scalability and performance. For large models, programming performance is better than Excel. The source code is publicly available at \textbf{\textit{Github}}\footnote{Source code: \url{https://github.com/Mahim-Sama/kmswc}}.

Composting systems must achieve two main objectives: producing high-quality compost efficiently and solving waste disposal issues. Key points include improving composting systems through optimizing ambient temperature, moisture levels, and aeration; producing high-quality compost free of contaminants; shortening composting time to make the process more efficient; and diverting organic waste from landfills to promote sustainable waste management practices. Overall, the emphasis is on advancing composting technologies to enhance sustainability in waste management and agriculture \citep{ref16}. The two-stage kinetic model of solid waste is part of the technological advancement of compost-related technology.

% ============================================================
\section{Methodology}

% [rev] "or stepped intermittent" airflow added.
The purpose of this project is to simulate and analyze the Kinetic Model of Solid Waste of Bari and Koenig (2012) \citep{ref17}. The project utilizes the Tkinter library to create a graphical user interface (GUI). Its main objective is to compute and visually represent parameters such as change in temperature, water content, volatile solids, and oxygen content due to continuous or stepped intermittent airflow over a specific period.

% [rev] Clarified what the global lists hold; "in the program" and "Each of them" added;
% [rev] layer geometry (6 layers of 0.2 m within a 1.2 m composting mass) added.
The code is organized into several sections: imports, global variables, GUI setup, main function, and helper functions. The imports section includes essential libraries, while the global variables section establishes the lists that store the hourly time series of each layer and the constants used throughout the program. There are two main functions in the program. Each of them handles all the calculations of 1\textsuperscript{st} stage composting and 2\textsuperscript{nd} stage composting consecutively. Loops are employed to simplify repetitive tasks (Algorithm~\ref{fig:loops}). Each of the composting models has 6 layers within a 1.2\,m composting mass, with each layer being 0.2\,m in height. There were 12 loops introduced for the 2-stage composting model. Nested loops are employed to iterate over the number of days and time steps, ensuring an accurate simulation of the system.

% \begin{figure}[ht]
%     \centering
%     \includegraphics[width=0.8\textwidth]{Figures/Figure_1.png}
%     \caption{Loops introduced for repetitive calculation.}
%     \label{fig:loops}
% \end{figure}

% (figure 1 png)
\begin{algorithm}[htbp]
\caption{Hourly heat and mass balance for one composting layer (repeated for 6 layers per stage).}
\label{fig:loops}
\KwIn{Initial layer mass fractions, ambient air properties, reaction constant $k$}
\KwOut{Time series of $T_a$, VS\%, WC\%, and $O_2$\% over the simulation period}
Initialize layer state ($m_{ai}$, $m_{ao}$, VS, $H_2O$, $T_a$, $kt$, WCf)\;
\For{$day \leftarrow 1$ \KwTo $28$}{
  \For{$hour \leftarrow 1$ \KwTo $24$}{
    $\Delta BVS \leftarrow WCf \cdot kt \cdot VS$ \tcp*{biodegradation this hour}
    $m_{ao} \leftarrow m_{ai} + 0.5\,\Delta BVS$ \tcp*{air mass balance}
    $\Delta H_2O \leftarrow H_2O_{out}\cdot m_{ao} - H_2O_{in}\cdot m_{ai}$\;
    Update VS, water mass, WC\%, and compost mass\;
    Compute thermal coefficient $K_L$ and heat terms $H_1\dots H_8$\;
    $T_a \leftarrow (H_1 + H_2 + H_3 + H_4 - H_5)\,/\,(H_6 + H_7 + H_8)$\;
    Compute $O_2\%$ (clamped at $0$)\;
    Recompute $kt$ and WCf for the next hour\;
  }
}
\end{algorithm}

The computed results are showcased in the GUI, presenting options for data visualization through graphs or charts. Matplotlib is leveraged for generating visualizations, allowing users to easily analyze and interpret the results. By simulating airflow and temperature distribution in a multi-layer system, the code delivers valuable insights and visualizations for users to interpret and analyze.

The GUI setup section is responsible for constructing the user interface, and the main function handles the primary calculations. The code imports Tkinter for GUI creation, the array library for array manipulations and NumPy for numerical calculations. The graphical user interface (GUI) is developed using Tkinter and includes entry widgets for user inputs such as temperature, relative humidity, different coefficients, and initial mass of solid waste, its water content, etc. Each input field is accompanied by a label and a button is available to initiate the main function. Furthermore, in the case of invalid input, an error message guides the user to rectify the input, thus safeguarding the program's stability by preventing crashes due to incorrect data.

% -------------------------------------------------------
\subsection{Flow Chart of Full Kinetic Model of Composting}

\begin{figure}[htbp]
    \centering
    \includegraphics[width=0.4\textwidth]{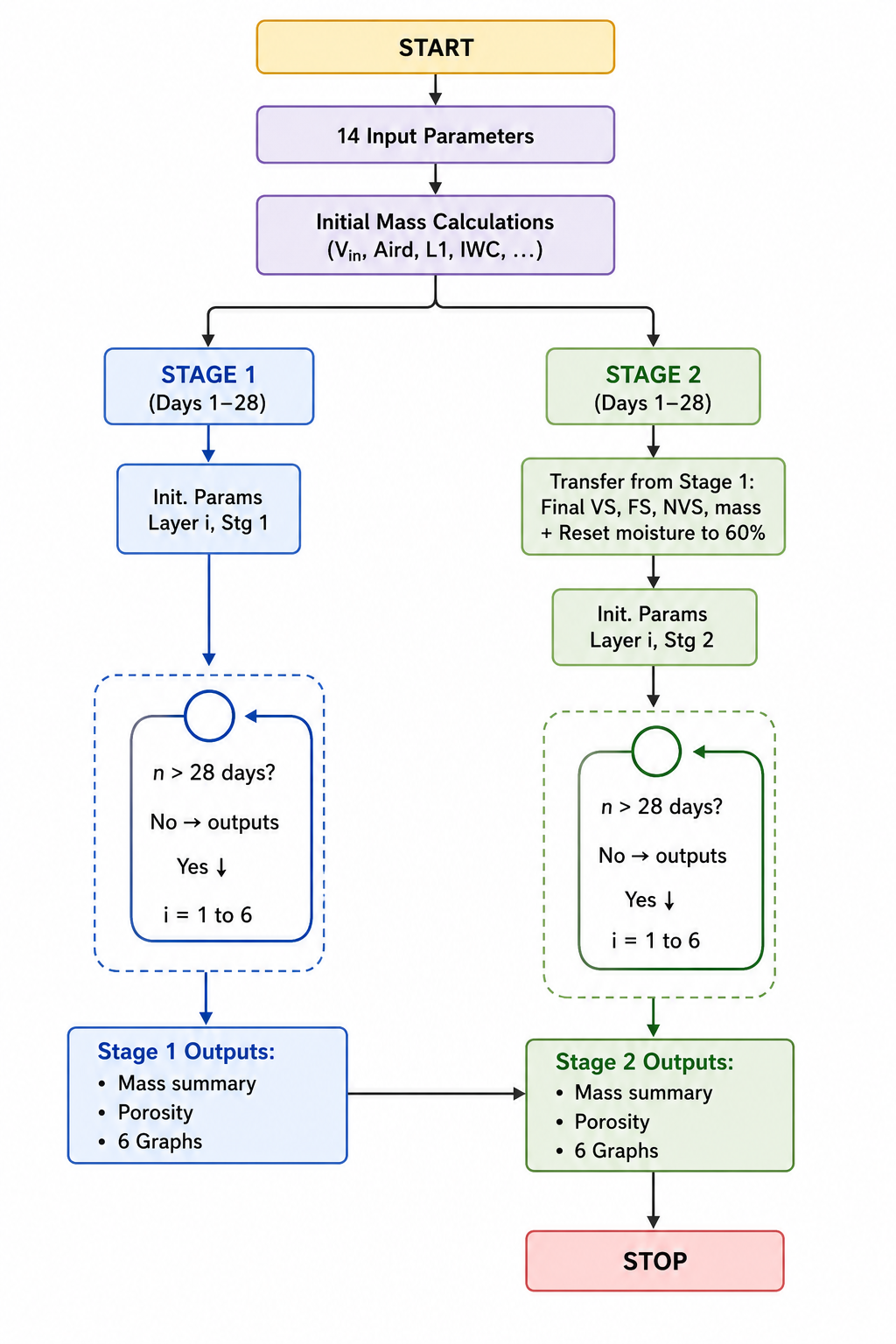}
    \caption{Flow chart of the programming-based kinetic model of composting.}
    \label{fig:flowchart}
\end{figure}

% [rev] Opening cross-reference sentence added.
The flow chart of the full kinetic model of composting is represented in Figure~\ref{fig:flowchart}. In the beginning of the program all the initial parameters are declared for input. Then necessary calculations are done before entering the loop. For the 6-layer mathematical model, a total of 6 loops were needed.

% Figure 2,3

% -------------------------------------------------------
\subsection{Physio-Chemical Parameters Calculation}

% Figure 2 (figure 4 png)
\begin{figure}[htbp]
\centering
\begin{lstlisting}[language=Python, basicstyle=\ttfamily\footnotesize, frame=single, breaklines=true]
# Input parameters
input_title = tk.Label(entry_frame, text="Input Parameters",
                       font=("Arial Bold", 14), fg="Navy Blue")
input_title.grid(row=0, column=0, columnspan=4, pady=win_pad, sticky="ew")

p_label1 = tk.Label(entry_frame, text="Initial Temperature of composting, Tci (C): ", font=win_font)
p_label1.grid(row=1, column=0, pady=win_pad, sticky="w")
p_entry1 = tk.Entry(entry_frame, font=win_font, width=10)
p_entry1.grid(row=1, column=1, pady=win_pad)
# ... remaining input fields follow the same Label / Entry / grid pattern
\end{lstlisting}
\caption{Python script for entry widget declaration.}
\label{fig:widgets}
\end{figure}

% [rev] Reaction rate (k) added to the input list so that the count of 14 is correct.
% [rev] The "single layer loop" sentence now points to Algorithm 1, not to the entry-widget figure.
There are a total of 14 inputs: initial temperature of composting mass ($T_c$), ambient temperature ($T_a$), heat energy generated by the degradation of BVS ($H_l$), specific heat capacity of wet composting material ($C_{pc}$), specific heat transfer coefficient ($K_c$), reaction rate ($k$), relative humidity (RH), initial mass of composting ($I\_mass$), initial percentage of water content ($P_{WC}$), initial percentage of fixed solid ($P_{FS}$), initial percentage of non-biodegradable volatile solids ($P_{NVS}$), initial percentage of volatile solids ($P_{VS}$), density of composting solid waste ($\rho$), and height of composting dumping ($h$). Figure~\ref{fig:widgets} shows the declaration of the entry widgets used to collect these inputs, and Algorithm~\ref{fig:loops} shows a single layer loop out of the 6 layers of a single-stage composting calculation for all the necessary parameters.
% \begin{figure}[ht]
%     \centering
%     \includegraphics[width=0.85\textwidth]{Figures/Figure_4.png}
%     \caption{Python script for entry widget declaration.}
%     \label{fig:widgets}
% \end{figure}

Weight of dry air ($Air_d$), mass of composting waste in Layer-1 ($L_1$), mass of initial water content ($I_{WC}$), mass of initial total solid ($I_{TS}$), mass of initial fixed solid ($I_{FS}$), mass of initial non-biodegradable volatile solid ($I_{NVS}$), mass of initial volatile solid ($I_{VS}$), mass of initial compost ($T_t$), and vapor inflow ($V_{in}$) are calculated using Equations~\ref{eq:vin}--\ref{eq:tt}:

\begin{equation}
    V_{in} = \frac{0.622 \times RH \times 10^{\left(8.896 - \frac{2238}{273 + T_a}\right)}}{760 - RH \times 10^{\left(8.896 - \frac{2238}{273 + T_a}\right)}}
    \label{eq:vin}
\end{equation}

\begin{equation}
    Air_d = 1.3009 - (0.0046 \times T_a)
    \label{eq:aird}
\end{equation}

\begin{equation}
    L_1 = \frac{I\_mass}{6}
    \label{eq:l1}
\end{equation}

\begin{equation}
    I_{WC} = \frac{P_{WC} \times L_1}{100}
    \label{eq:iwc}
\end{equation}

\begin{equation}
    I_{TS} = L_1 - I_{WC}
    \label{eq:its}
\end{equation}

\begin{equation}
    I_{FS} = \frac{I_{TS} \times P_{FS}}{100}
    \label{eq:ifs}
\end{equation}

\begin{equation}
    I_{NVS} = \frac{I_{TS} \times P_{NVS}}{100}
    \label{eq:invs}
\end{equation}

\begin{equation}
    I_{VS} = \frac{I_{TS} \times P_{VS}}{100}
    \label{eq:ivs}
\end{equation}

\begin{equation}
    T_t = I_{WC} + I_{FS} + I_{NVS} + I_{VS}
    \label{eq:tt}
\end{equation}

% -------------------------------------------------------
\subsection{Second Stage Composting}

% [rev] "to the dry mass of the first stage compost" added.
After 28 days, the first stage is done. But for further degradation, an extra amount of water needs to be added to the dry mass of the first stage compost. Because of lower moisture content, microbial activity slows down. To show further degradation by increasing moisture to the optimum level of around 60\% \citep{ref18} and other initial parameters, the second step of composting mass can be done in this programming-based model. The final mass of volatile solid and final height of the composting pile will be the initial mass of volatile solid and initial height of the second step composting.

The two-stage process produces more mature and stable compost, reducing the risk of pathogens and harmful compounds while improving soil enrichment \citep{ref19}. It also reduces the total composting time compared to traditional methods \citep{ref20}.

% ============================================================
\section{Results and Discussion}

\subsection{General}

Using the programming-based kinetic model of solid waste composting, all calculations, graphical representations, and output tables are generated. The input parameter coefficients are kept the same to compare among the graphs.

% -------------------------------------------------------
% [rev] Subsection title updated to match the revised manuscript.
\subsection{Comparison Between Previously Developed Kinetic Model of Solid Waste Composting and the Present Programming-Based Model}

For the same input parameters, a comparison between the previously made Excel model and this programming-based model is presented.

\begin{figure}[htbp]
    \centering
    \includegraphics[width=0.7\textwidth]{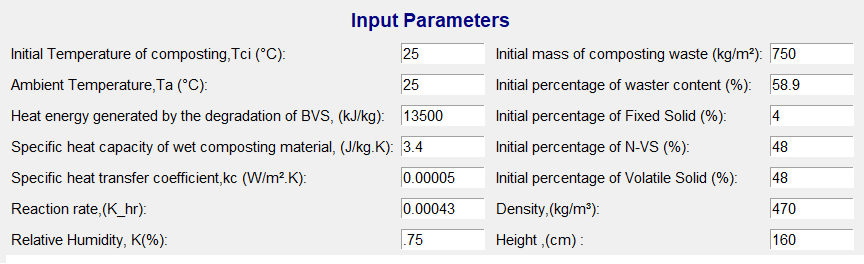}
    \caption{Set input parameters in the software.}
    \label{fig:input_params}
\end{figure}

% [rev] Units added to HL, kc, k and the initial mass, per the revision.
The input parameters used are: Initial Temperature of composting, $T_{ci} = 25$\,\textdegree{}C; Ambient Temperature, $T_a = 25$\,\textdegree{}C; Heat energy generated by the degradation of BVS, $H_L = 13500$\,kJ/kg; Specific heat capacity of wet composting material, $c_{pc} = 3.4$; Specific heat transfer coefficient, $k_c = 0.00005$\,W\,m\textsuperscript{-2}\,K\textsuperscript{-1}; Reaction rate, $k = 0.00043$\,h\textsuperscript{-1}; Relative Humidity, $RH = 0.75$; Initial mass of composting waste $= 750$\,kg\,m\textsuperscript{-2}; Initial water content $= 58.9$\%; Initial Fixed Solid $= 4$\%; Initial Non-Biodegradable Volatile Solid $= 48$\%; Initial Volatile Solid $= 48$\%; Density $= 470$\,kg/m\textsuperscript{3}; Height $= 160$\,cm; and a constant air flow of 5\,m\textsuperscript{3}\,m\textsuperscript{-2}\,h\textsuperscript{-1} is applied for 28 days continuously. Figure~\ref{fig:input_params} shows a sample of how to import the parameters.

% Figure 3
% [rev] NOTE: the revised manuscript asks for the screenshot itself to be corrected --
% [rev] the reaction rate label should read "k" (not "K") and the relative humidity
% [rev] label should read "RH" (not "K"). Regenerate Figure 5.png before submission.

\begin{figure}[htbp]
    \centering
    \includegraphics[width=0.85\textwidth]{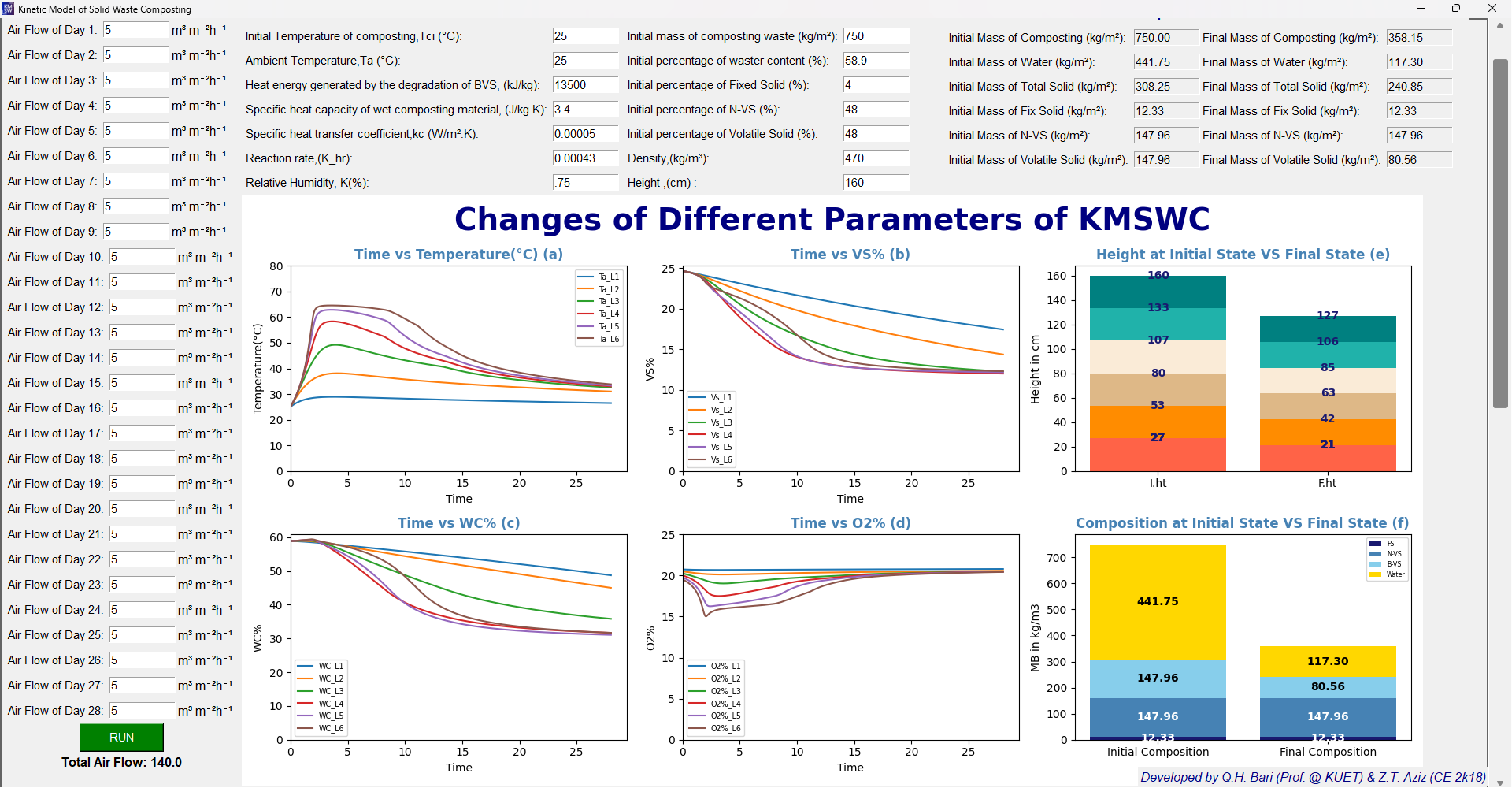}
    \caption{Change in different parameters of composting process at $T_a = 25$\,\textdegree{}C and $Q_i = 5.0$\,m\textsuperscript{3}\,m\textsuperscript{-2}\,h\textsuperscript{-1} developed using programming.}
    \label{fig:prog_graphs}
\end{figure}

\begin{figure}[htbp]
    \centering
    \includegraphics[width=0.75\textwidth]{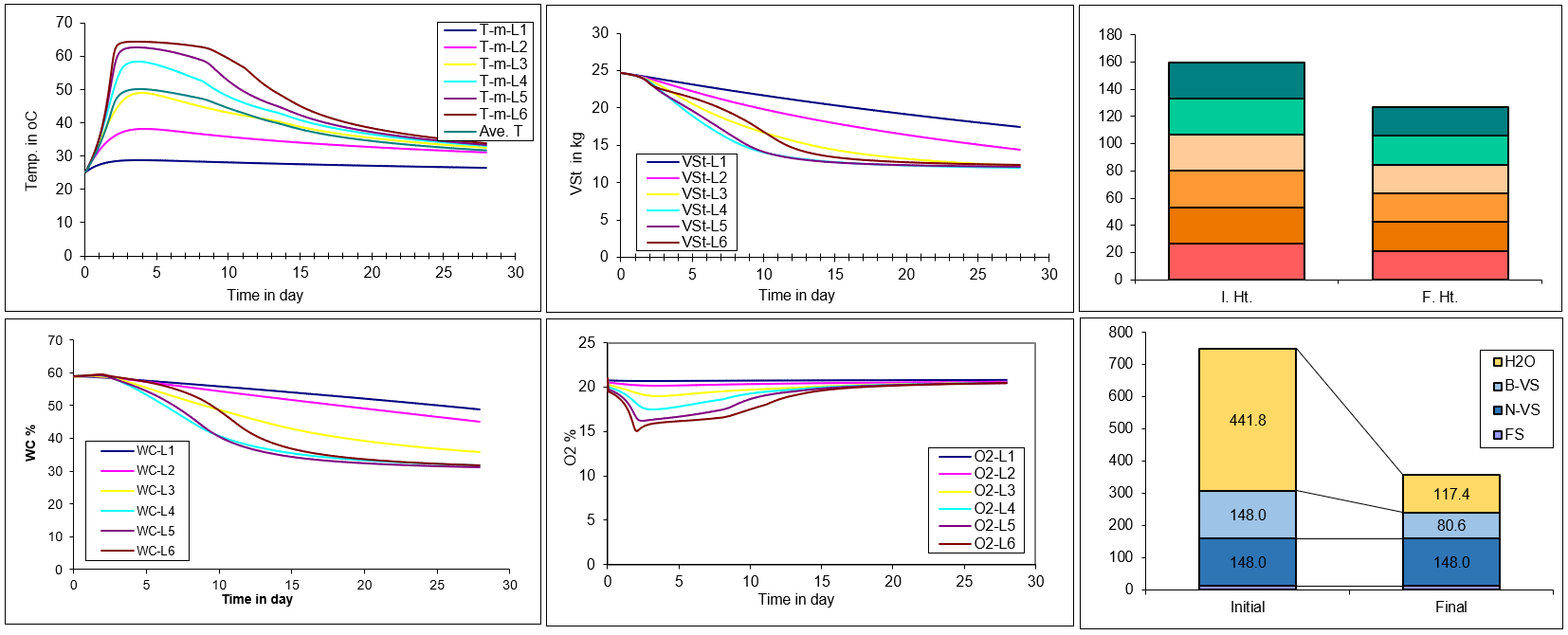}
    \caption{Change in different parameters of composting process at $T_a = 25$\,\textdegree{}C and $Q_i = 5.0$\,m\textsuperscript{3}\,m\textsuperscript{-2}\,h\textsuperscript{-1} using MS Excel Spreadsheet \citep{ref17}.}
    \label{fig:excel_graphs}
\end{figure}

% [rev] "water content"; "the same" -> "similar"; "sheet model" -> "spreadsheet model".
In Figure~\ref{fig:prog_graphs}, all the graphs of change in different parameters of the composting process are developed using coding. They are exactly the same as Figure~\ref{fig:excel_graphs}, which was developed previously by Excel, a mathematical model of composting of solid waste \citep{ref11, ref17}. Initial and final values of water content, volatile solid, non-biodegradable volatile solid, and fixed solid are similar for both the programming-based model and the previously developed spreadsheet model. So, the programming-based model of composting of solid waste is reliable and the further findings using this model will be the same as the previously proved composting model calculated using Excel spreadsheet.

% Figure 4

% Figure 5

For both Figure~\ref{fig:prog_graphs} and Figure~\ref{fig:excel_graphs}, the input parameters are the same, and for the same input parameters both models give the same output.

% Figure 6
\begin{figure}[htbp]
    \centering
    \includegraphics[width=0.6\textwidth]{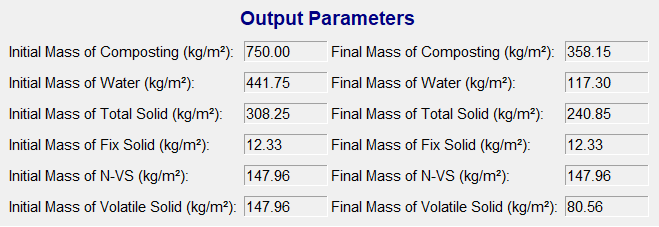}
    \caption{Initial and final condition of composted solid waste developed using the software.}
    \label{fig:initial_final_s1}
\end{figure}

In Figure~\ref{fig:initial_final_s1}, a comparison between initial mass and final mass of composted solid waste, and the difference between different composting parameters, are shown.

% -------------------------------------------------------
\subsection{Second Stage Composting}

% [rev] "microbial activities" removed from the list of variations analysed.
A comparison is made between 1\textsuperscript{st} stage composting and 2\textsuperscript{nd} stage composting. Variation of temperature and volatile solid degradation rate are analyzed to determine how much degradation can occur after 1\textsuperscript{st} stage composting. 1\textsuperscript{st} stage composting is done for 28 days with constant air flow of 5\,m\textsuperscript{3}\,m\textsuperscript{-2}\,h\textsuperscript{-1}.

% Figure 7
\begin{figure}[htbp]
    \centering
    \includegraphics[width=0.7\textwidth]{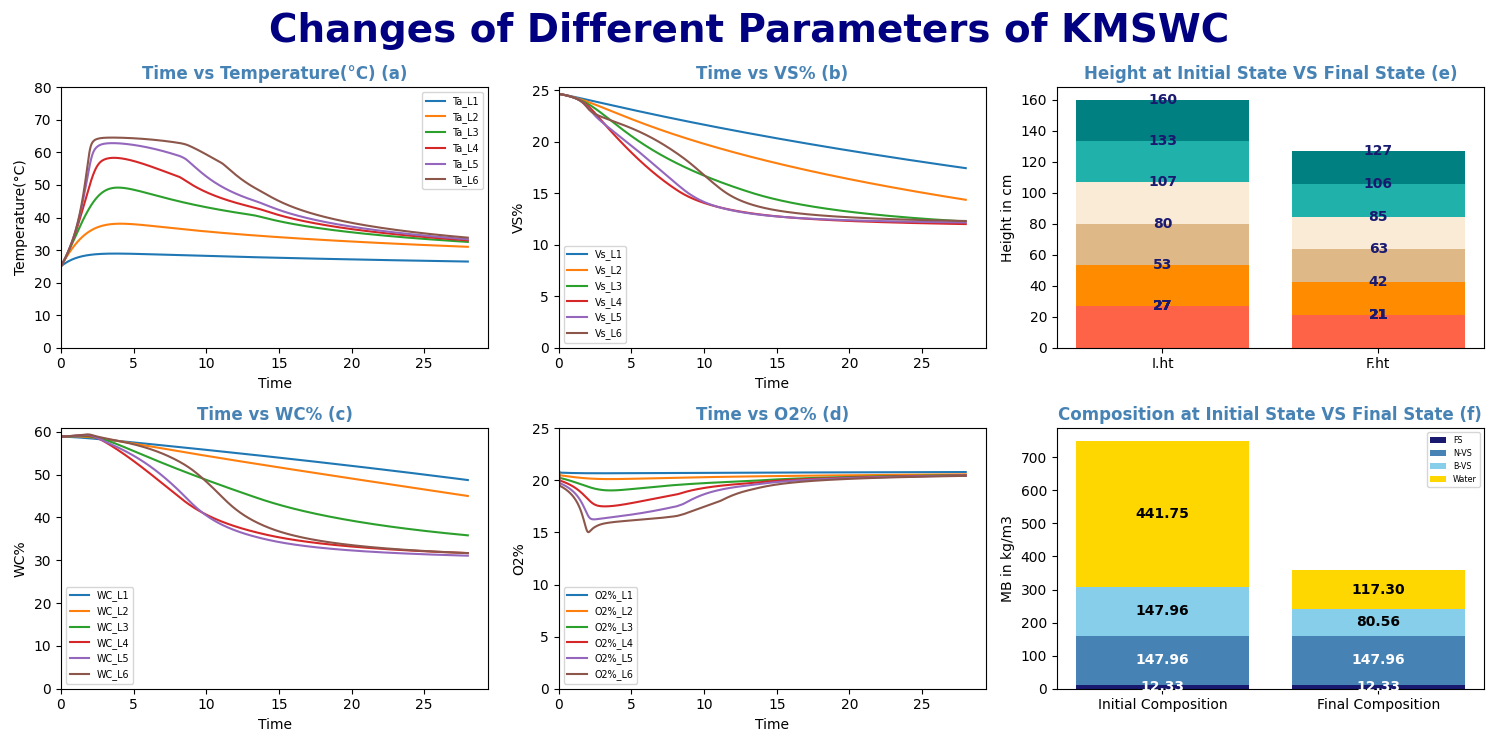}
    \caption{Change in different parameters of composting process for 1\textsuperscript{st} stage composting.}
    \label{fig:stage1_graphs}
\end{figure}

% [rev] "percentage" deleted; "data of" added.
After 28 days of composting, the waste turns into a solid humus-like material and becomes dry \citep{ref21}. At that condition it will not further degrade. But with the addition of moisture content, the mass can be further degraded. For second step composting, the moisture content is raised to 60\%. The data of initial mass of fixed solid, volatile solid, and non-biodegradable volatile solid are taken from the final values of 1\textsuperscript{st} stage composting. The software will automatically take the initial mass of solid waste composting, fixed solid, non-volatile solid, and volatile solid from the final output of stage 1.

% Figure 8
\begin{figure}[htbp]
    \centering
    \includegraphics[width=0.7\textwidth]{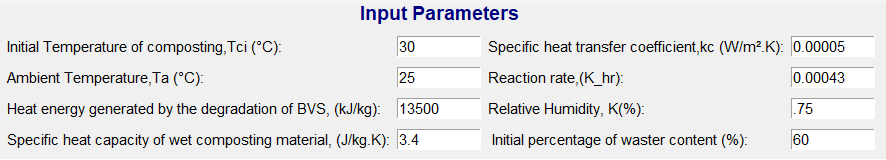}
    \caption{Input parameters for 2\textsuperscript{nd} stage composting.}
    \label{fig:stage2_input}
\end{figure}

% Figure 9
\begin{figure}[htbp]
    \centering
    \includegraphics[width=0.7\textwidth]{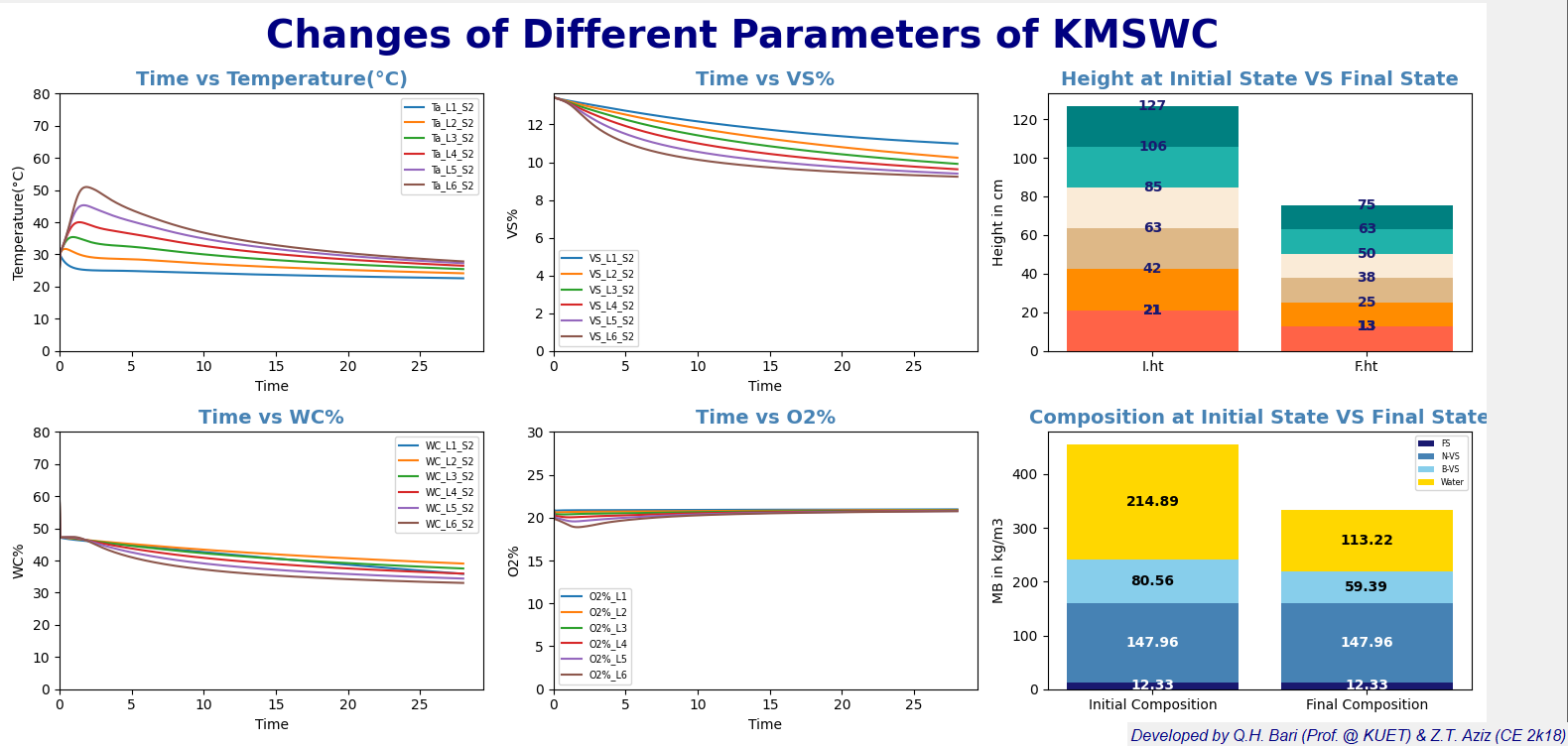}
    \caption{Change in different parameters of composting process for 2\textsuperscript{nd} stage composting.}
    \label{fig:stage2_graphs}
\end{figure}

% [rev] "temperature variation is not seen much" -> "temperature distribution in a narrow band
% [rev] is observed"; the layer-wise rise is now quantified as "up to 50 C".
After analyzing the physio-chemical parameters of the 2\textsuperscript{nd} stage composting, different deductions can be made. In the Time vs.\ Temperature graph, a temperature distribution within a narrow band is observed, as seen in Figure~\ref{fig:stage2_graphs}. The temperature of the bottom layer is around 25\textdegree{}C. As the layer increases, the temperature slightly increases, up to 50\textdegree{}C. After 4--5 days, no significant variation in temperature is seen. Temperature below 40\textdegree{}C means microbial activity is at a minimum level \citep{ref18, ref21}. In the Time vs.\ VS\% graph, the change in volatile solid is very low compared to the 1\textsuperscript{st} stage. The overall reduction of VS is 10\% for 1\textsuperscript{st} stage and 2\% for 2\textsuperscript{nd} stage. Less variation of volatile solid means less microbial activities. In the Time vs.\ O$_2$\% graph, the average value of O$_2$ is around 21\%. At this stage, microbial activities are not usually observed. The height difference of the composting mass is mainly seen due to reduction of moisture content. This means at this stage the maximum portion of water has vaporized. From Figure~\ref{fig:stage1_graphs} and Figure~\ref{fig:stage2_graphs}, we can compare changes of mass of different parameters of the composted sample.

% Figure 10
\begin{figure}[H]
    \centering
    \includegraphics[width=0.55\textwidth]{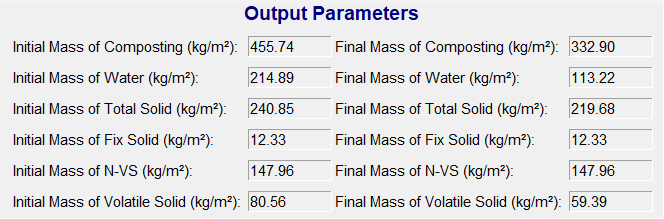}
    \caption{Initial and final condition of composted solid waste developed using programming after 2\textsuperscript{nd} stage composting.}
    \label{fig:final_stage2}
\end{figure}

% [rev] The revised manuscript flags this sentence with "Clarify the figure" --
% [rev] the description below should be expanded once the intended clarification is decided.
After 2\textsuperscript{nd} stage composting, the final output is shown in Figure~\ref{fig:final_stage2}.

% ============================================================
\section{Conclusions}

The development and implementation of a programming-based kinetic model for two-stage solid waste composting is a significant step forward in sustainable waste management technology. This study presents a systematic method to understanding and optimizing the composting process by combining mathematical modeling and computer tools.

The kinetic model developed here gives important insights into the complicated biochemical and physicochemical processes that occur during composting. The model makes it easier to build and run effective composting systems by reliably forecasting critical factors including temperature, moisture content, percentage of volatile solid, and oxygen consumption rate. A programming-based model of composting is cost-effective, less time consuming, easy to understand, simple, and a user-friendly approach for waste management.

Second stage composting analysis provides information about how much solid waste can go through further degradation.

The programming-based kinetic model of solid waste composting will simplify further research in composting and waste management, as the experimental approach for composting is very sophisticated, time consuming, and expensive.

% ============================================================
\bibliographystyle{unsrtnat}
\bibliography{references}

\end{document}